\documentclass[aps,prl,amsmath,amssymb,amstext,citeautoscript,punctuation,nofootinbib,superscriptaddress,twocolumn]{revtex4-1}
\usepackage[dvips]{graphicx}
\usepackage{amsmath,amssymb,amsthm,mathrsfs,amsfonts,dsfont,hyperref,subfigure, epsfig, braket,bm,enumerate,color,graphicx,dcolumn}
\usepackage[normalem]{ulem}
\usepackage[dvipsnames]{xcolor}
\usepackage{{xcolor,charter, gensymb, float}}
\usepackage [english]{babel}

\newcommand{\eaa}{EuAgAs}
\begin{document}
\title{Higher-order Weyl nodes driven by helical magnetic order in EuAgAs}
\author{Jian-Rui Soh}
\affiliation{Quantum Innovation Centre (Q.InC), Agency for Science, Technology and Research (A*STAR), 2 Fusionopolis Way, Innovis \#08-03, Singapore 138634, Singapore}
\affiliation{Centre for Quantum Technologies, National University of Singapore, 3 Science Drive 2, Singapore 117543, Singapore}
\author{Ziming Zhu}
\affiliation{School of Physics and Electronics, Hunan Normal University, Key Laboratory for Matter Microstructure and Function of Hunan Province,Key Laboratory of Low-Dimensional Quantum Structures and Quantum Control of Ministry of Education, Changsha 410081, China}
\author{Louis Withers}
\affiliation{Department of Physics, University of Oxford, Clarendon Laboratory, Oxford OX1 3PU, United Kingdom}
\author{J. Alberto Rodr\'iguez-Velamaz\'an}
\affiliation{Institut Laue-Langevin, 6 rue Jules Horowitz, BP 156, 38042 Grenoble Cedex 9, France}
\author{Timur K. Kim}
\affiliation{Diamond Light Source, Harwell Science and Innovation Campus, Didcot, Oxfordshire OX11 0DE, United Kingdom}
\author{Oscar Fabelo}
\affiliation{Institut Laue-Langevin, 6 rue Jules Horowitz, BP 156, 38042 Grenoble Cedex 9, France}
\author{Anne Stunault}
\affiliation{Institut Laue-Langevin, 6 rue Jules Horowitz, BP 156, 38042 Grenoble Cedex 9, France}
\author{Daniil Yevtushynsky}
\affiliation{Laboratory for Quantum Magnetism, Institute of Physics, \'Ecole Polytechnique F\'ed\'erale de Lausanne, CH-1015 Lausanne, Switzerland}
\author{Dharmalingam Prabhakaran}
\affiliation{Department of Physics, University of Oxford, Clarendon Laboratory, Oxford OX1 3PU, United Kingdom}
\author{Shengyuan A. Yang}
\affiliation{Institute of Applied Physics and Materials Engineering, Faculty of Science and Technology, University of Macau, Macau, China}
\author{Andrew T. Boothroyd}
\affiliation{Department of Physics, University of Oxford, Clarendon Laboratory, Oxford OX1 3PU, United Kingdom}
\date{\today}
\begin{abstract}
Magnetic topological semimetals provide a fertile ground for exploring how long-range magnetic order can alter electronic band structures and generate {\color{black}novel} quasiparticles such as Weyl fermions. Here, we investigate the coupled magnetic and electronic structure of single-crystalline EuAgAs, a hexagonal pnictide whose magnetic ground state has remained elusive. Using neutron diffraction and resonant elastic X-ray scattering, we identify an unusual magnetic ordering sequence with two successive phase transitions at $T_\mathrm{N1} = 12$\,K and $T_\mathrm{N2} = 8$\,K. {\color{black}We observe two slightly different magnetic  propagation vectors, one associated with $T_\mathrm{N1}$ and the other with  $T_\mathrm{N2}$. Spherical neutron polarimetry reveals that the magnetic structure is a transverse helix aligned along the $c$ axis with a period that is approximately twice the $c$ lattice parameter.} 
{\color{black}First-principles calculations for the helical phase predict subtle band folding effects and the emergence of effective higher-order Weyl nodes. These topological features appear near the calculated Fermi energy $E_{\mathrm{F}}$ which, however, lies above the position of $E_{\mathrm{F}}$ obtained from angle-resolved photoemission spectroscopy so could not be probed in this study. }
\end{abstract}
\maketitle
\section{Introduction}

The interplay between magnetism and band topology has become a central theme in condensed matter physics, driven by the quest to realise novel quantum states and exotic quasiparticles such as magnetic Weyl and Dirac fermions~\cite{Armitage2018, Burkov2016,Boothroyd02102022}. In topological semimetals, the breaking of time-reversal by magnetic order can lift degeneracies in the electronic band structure, giving rise to Weyl points, nodal lines, or higher-order topological features that are protected by crystalline symmetries~\cite{Wan2011, Soh_2023_EIA, Liu2019}. Such systems are of considerable interest not only for their fundamental physics but also for their potential applications in spintronics and topological devices~\cite{Ma_Review_2025}.

\begin{figure}[b!]
\includegraphics[width=0.49\textwidth]{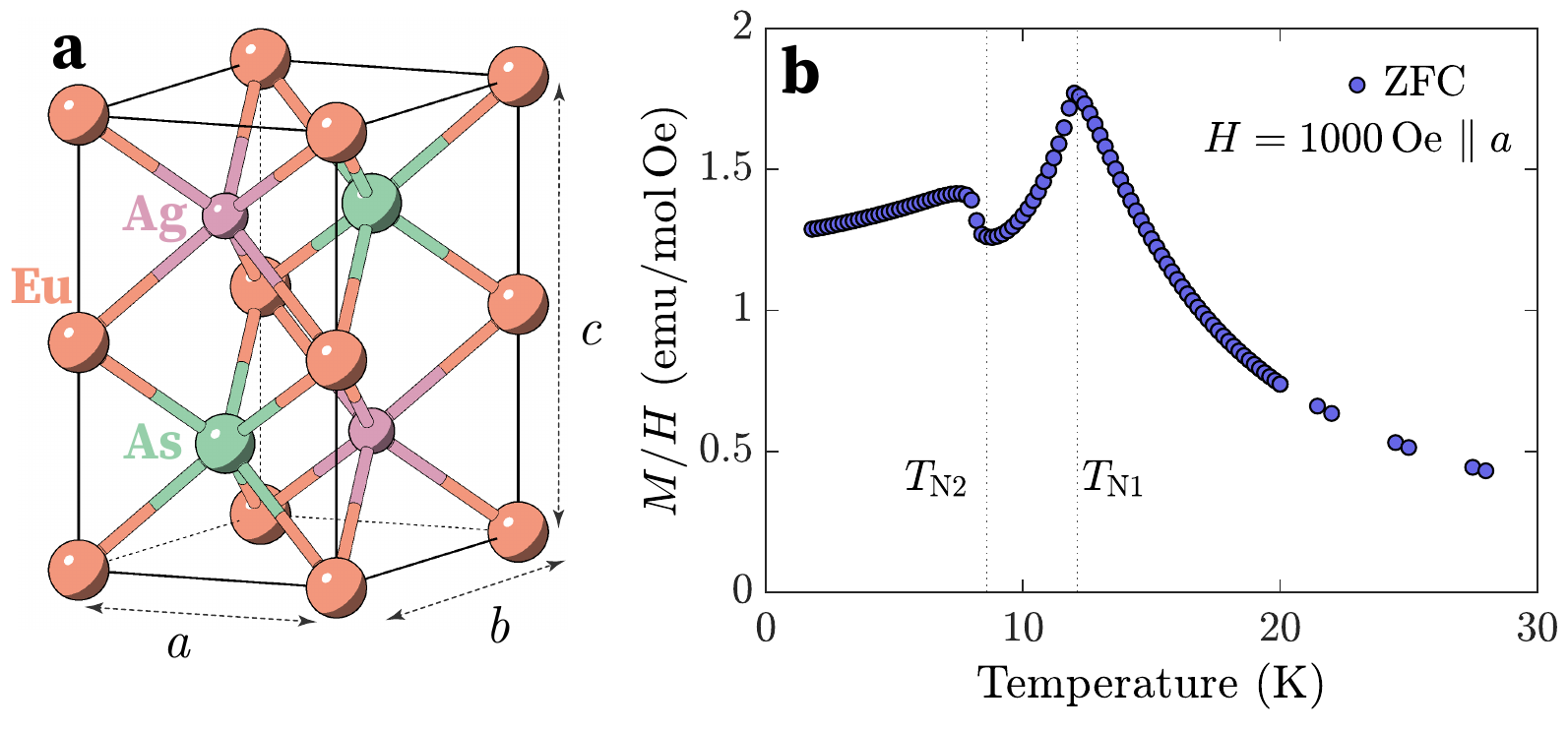}
\caption{\label{fig:Figure_1} \textbf{a} The hexagonal unit cell of \eaa{} can be described by the $P6_3/mmc$ space group. 
\textbf{b} The magnetic susceptibility of \eaa{} with the magnetic field ($\textbf{H}$) along the crystal $a$ axis displays two anomalies, at $T_\mathrm{N1}$=12\,K and $T_\mathrm{N2} \simeq 8$\,K.}
\end{figure}

Rare-earth intermetallic compounds, in particular, have emerged as a rich class of materials in which the localized $4f$ moments couple to itinerant electrons, enabling tunable topological phases via magnetic ordering and applied field \cite{Soh_2023_EIA, Soh_2024_EuCuAs, roychowdhury_interplay_2023}. A prototypical example is EuCuAs, in which a helical spin structure induces a doubling of the magnetic unit cell and generates higher-order Weyl nodes in the electronic spectrum \cite{Soh_2024_EuCuAs}. However, direct experimental confirmation of similar magnetic and electronic behavior in structurally related compounds has remained scarce, {\color{black}making it difficult to assess how widely applicable this mechanism is}.

\begin{figure*}[ht!]
\includegraphics[width=0.65\textwidth]{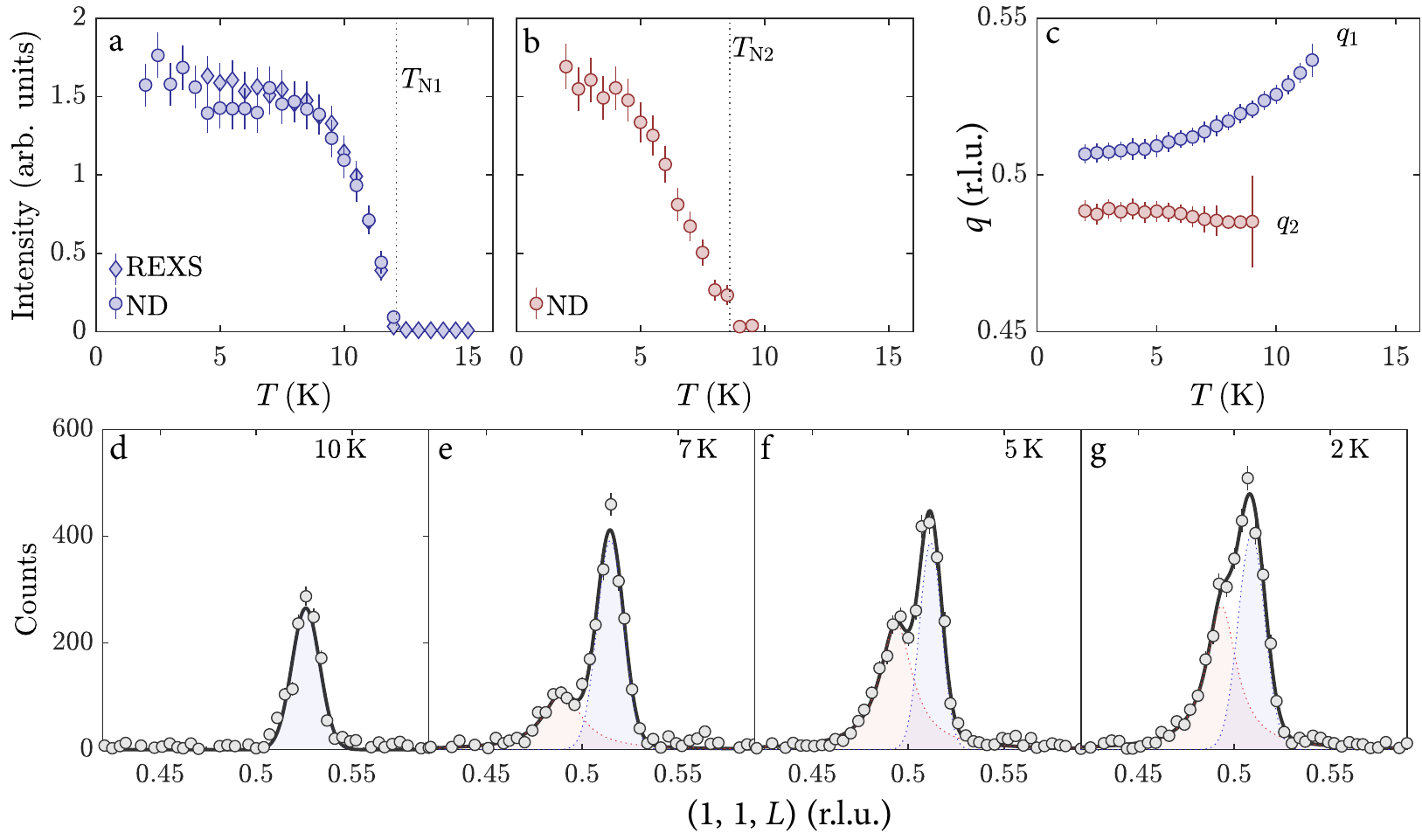}
\caption{\label{fig:Figure_2} \textbf{Unravelling the magnetic order of \eaa{} with neutrons and x-rays.} \textbf{a} Temperature dependence of the \textbf{q}$_1$ peak intensity obtained with resonant elastic x-ray scattering (REXS) and neutron diffraction (ND). \textbf{b}. Intensity of the $\mathbf{q}_2$ magnetic reflection measured with ND. \textbf{c} Temperature evolution of the reciprocal space location of the $q_1$ and $q_2$ reflections. 
\textbf{d}--\textbf{g} Scans along the line $(1,1,L)$ in reciprocal space  at $T$\,=\,10, 7, 5 and 2\,K. The lines are fits to Gaussian ($\mathbf{q}_1$) and Lorentzian ($\mathbf{q}_2$) peak functions. The error bars in \textbf{a}--\textbf{c} correspond to the standard deviations on the peak fit parameters, and the errors in  \textbf{d}--\textbf{g} are from counting statistics. 
}
\end{figure*}
Here, we focus on EuAgAs, a hexagonal pnictide that shares the same crystal structure as EuCuAs (see Fig.~\ref{fig:Figure_1}a)  and which is reported to exhibit a sizable topological Hall effect~\cite{EuAgAs_DFT_Magnetization}. A range of different topological bulk and surface electronic states has been predicted depending on the type of magnetic order that forms at temperatures below 12\,K~\cite{Jin_EuAgAs_DFT}. Previous  studies~\cite{EuAgAs_DFT_Magnetization, Jin_EuAgAs_DFT, Zhang_EuAgAs_2023, liu_arXiv_2024} have proposed that the Eu moments {\color{black}in EuAgAs} align within the basal plane to form ferromagnetic layers that stack antiferromagnetically along the crystal $c$ axis. In this collinear spin configuration, the magnetic unit cell coincides with the structural unit cell of the hexagonal $P6_3/mmc$ space group. However, direct experimental evidence for this assumed magnetic arrangement is lacking,  and so the effect of magnetic order on the electronic structure cannot be confirmed. 

In this work, we combine unpolarised neutron diffraction, resonant elastic X-ray scattering, spherical neutron polarimetry, angle-resolved photoemission spectroscopy (ARPES) with first-principles calculations to clarify the magnetic ground state and its consequences for the low-energy electronic states in EuAgAs. Our measurements uncover a non-trivial helical spin configuration that roughly doubles the unit cell along the $c$ axis. ARPES spectra reveal that the Fermi surface remains largely unaffected by the onset of helical order. The ARPES results are in agreement with density functional theory (DFT) calculations, with applied downward shift of the calculated Fermi energy by approximately 0.8 eV. 
Near the calculated $E_\mathrm{F}$, DFT predicts subtle band folding and the emergence of Weyl points protected by the helical magnetic order.

Our findings position EuAgAs as a promising magnetic topological semimetal with a higher order Weyl node generated by the helical spin configuration, providing an ideal platform for future studies of emergent Weyl physics driven by non-collinear magnetism.

\section{Results}
\subsection{Magnetic susceptibility}

Figure~\ref{fig:Figure_1}\textbf{b} shows the magnetic susceptibility of \eaa{} measured with the external magnetic field applied along the crystal $a$ axis after cooling in zero field. The magnetic susceptibility exhibits two successive anomalies, one at $T_\mathrm{N1} = 12$\,K and the second at $T_\mathrm{N2} \simeq 8$\,K. These features are consistent with earlier measurements~\cite{EuAgAs_DFT_Magnetization,Zhang_EuAgAs_2023, liu_arXiv_2024}.

\subsection{Magnetic diffraction with x-rays and neutrons}
To elucidate the evolution of the magnetic configuration across these two transitions, we performed unpolarised neutron diffraction (ND) and resonant elastic x-ray scattering (REXS) on single-crystalline \eaa{}. Both techniques reveal the emergence of magnetic Bragg peaks signaling magnetic order below $T_\mathrm{N1} = 12$\,K, coinciding with the onset of the first anomaly observed in the magnetic susceptibility [Fig.~\ref{fig:Figure_1}\textbf{b}].  The intensity of the magnetic peaks follows an order parameter-like temperature dependence, as shown in Fig.~\ref{fig:Figure_2}\textbf{a}.

\begin{figure}[t!]
\includegraphics[width=0.45\textwidth]{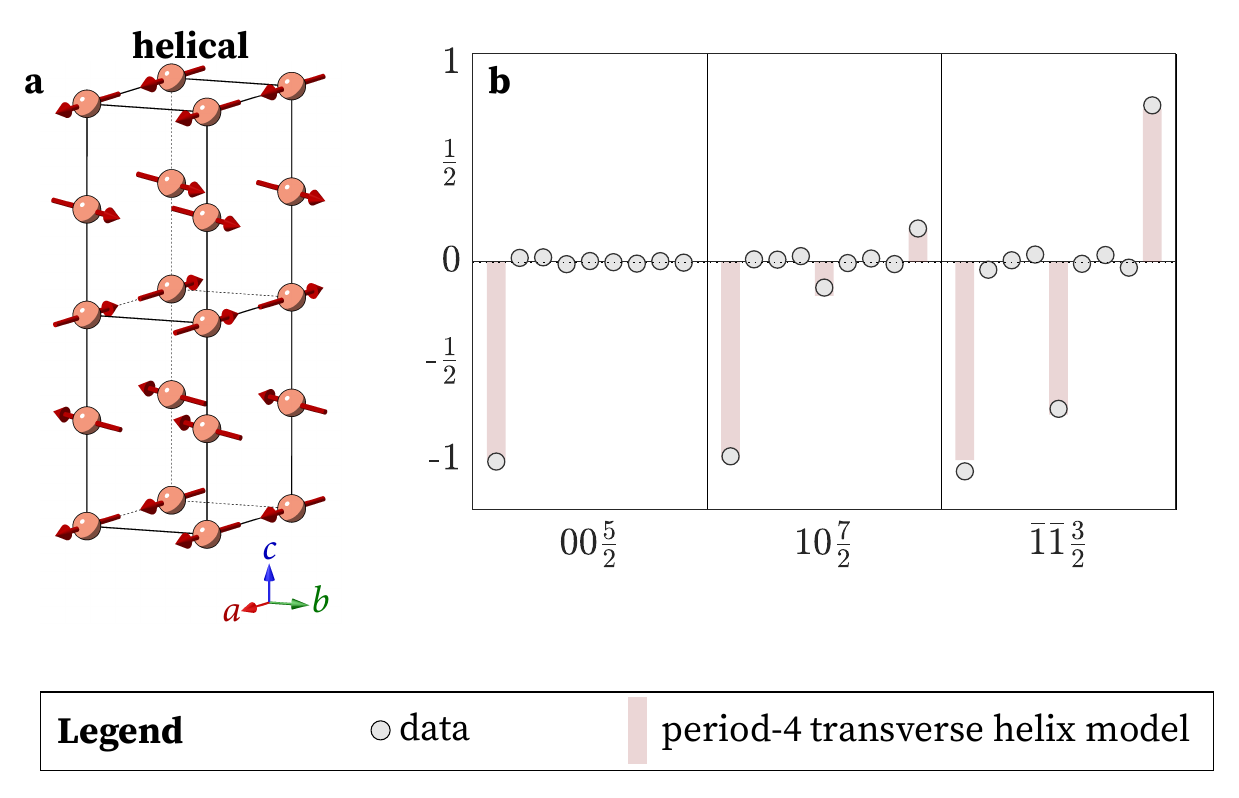}
\caption{\label{fig:Figure_3} \textbf{Determining the ground state magnetic order of \eaa{} using spherical neutron polarimetry.} {\color{black}\textbf{a} Helical magnetic structure with unit cell doubling along the crystal $c$ axis.  \textbf{b} The grey filled circles show the measured polarization matrices ($P_{ij}^\mathrm{obs.}$) for three magnetic reflections at $T = 2$\,K. For each reflection, the nine elements of the polarization matrix $P_{ij}$ from left to right correspond to $ij = xx$, $xy$, $xz$, $yx$, $yy$, $yz$, $zx$, $zy$, and $zz$. The red bars indicate the calculated polarization matrices ($P_{ij}^\mathrm{calc.}$) assuming a period-4 helical magnetic structure.}}
\end{figure}

The associated Eu spin configuration can be described by the magnetic propagation vector $\mathbf{q}_1$=$(0,0,0.5+\delta_1)$. Interestingly, the reciprocal space position of the corresponding magnetic peak shifts with temperature [Fig.~\ref{fig:Figure_2}\textbf{c}]. On cooling from $T_\mathrm{N1}$, the magnitude of $\delta_1$ decreases continuously from $\delta_1 = 0.04$ at 12\,K to $\delta_1 = 0.01$ at 2\,K, indicating a gradual evolution of the size of the magnetic supercell.
    
In addition, ND reveals a second magnetic peak with a different propagation vector, $\mathbf{q}_2$=$(0,0,0.5-\delta_2)$, which emerges below $T_\mathrm{N2} = 8$\,K, as shown in Figs.~\ref{fig:Figure_2}\textbf{b}, \textbf{c}. In contrast to $\delta_1$, the magnitude of $\delta_2$ remains fairly constant at 0.01 down to the lowest temperatures measured.

Figure~\ref{fig:Figure_2}\textbf{d}--\textbf{g} plots scans along the $(1,1,L)$ reciprocal space direction across the two magnetic peaks, obtained at various temperatures. Between $T_\mathrm{N1}$ and $T_\mathrm{N2}$ (Fig.~\ref{fig:Figure_2}\textbf{d}), only the $\mathbf{q}_1 = (0,0,0.5+\delta_1)$ reflection is present. Upon cooling below $T_\mathrm{N2}$ (Figs.~\ref{fig:Figure_2}\textbf{e}, \textbf{f}), the second peak, $\mathbf{q}_2 = (0,0,0.5-\delta_2)$, appears as a shoulder to the $\mathbf{q}_1$ reflection, demonstrating the coexistence of both magnetic modulations. On further cooling, the two peaks move closer together and eventually become unresolved at $T = 2$\,K. The resulting lineshape is centred on $(0,0,0.5)$, as seen in Fig.~\ref{fig:Figure_2}\textbf{g}.

The magnetic periodicity observed here by diffraction cannot be reconciled with the A-type antiferromagnetic (AFM) structure assumed in previous studies of \eaa{}~\cite{EuAgAs_DFT_Magnetization,Zhang_EuAgAs_2023,liu_arXiv_2024,Jin_EuAgAs_DFT}. In the latter, the magnetic unit cell coincides with the crystallographic unit cell, corresponding to a propagation vector $\mathbf{q}_\mathrm{AF} = (0, 0, 1)$. Instead, our ND results demonstrate that at $T = 2$\,K the magnetic supercell is approximately doubled along the $c$ axis relative to the structural unit cell.


\begin{figure}[t!]
\includegraphics[width=0.48\textwidth]{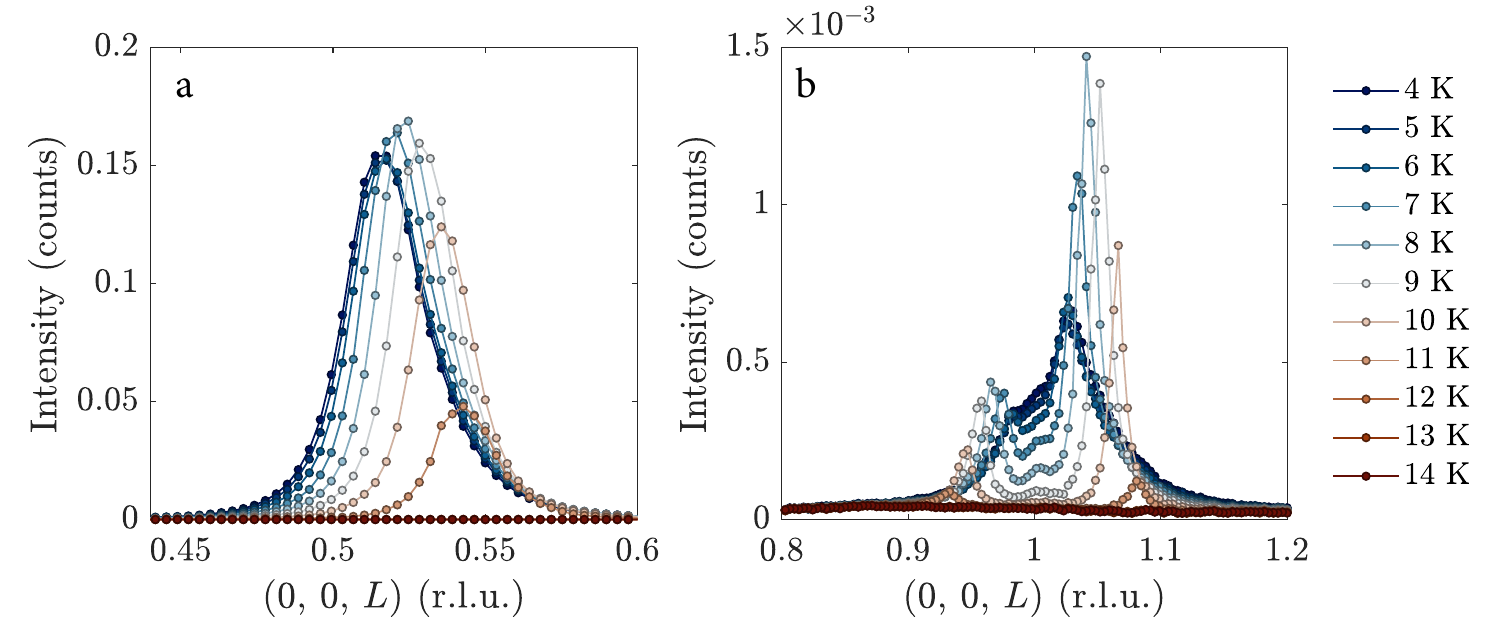}
\caption{\label{fig:Figure_REXS} \textbf{First and second harmonics of the $\mathbf{q}_1$ magnetic peak. }  The scans are along the $(0,0,L)$ direction in reciprocal space, and were measured with x-rays tuned to the Eu $M_5$ absorption edge in the $\pi \rightarrow \sigma'$ polarization channel. \textbf{a} The $\mathbf{q}_1 = (0,0,0.5+\delta_1)$ magnetic peak. \textbf{b}  Second harmonics of the $\mathbf{q}_1$ peak.} 
\end{figure}

\begin{figure*}[ht!]
\includegraphics[width=0.89\textwidth]{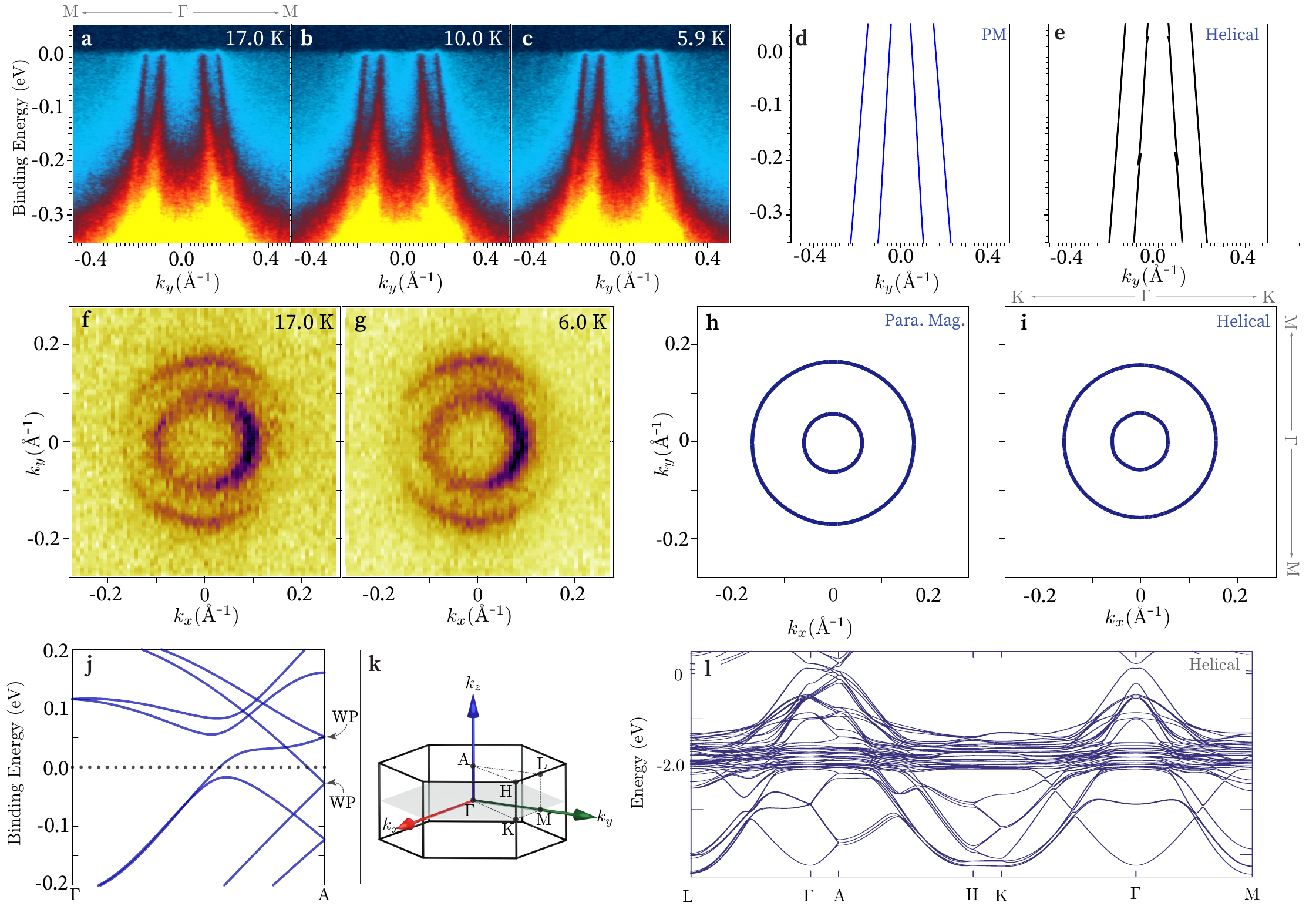}
\caption{\label{fig:Figure_4} \textbf{Comparison between the measured and calculated electronic band structure of \eaa{}.} \textbf{a}-\textbf{c} Experimental dispersion along M$-$$\Gamma$$-$M at three temperatures ($T =17$, 10 and 5.9\,K). \textbf{d}, \textbf{e} Calculated dispersion along high symmetry lines in the paramagnetic and helical magnetic phase. An upward shift of 0.5\,eV has been applied to the calculated bands shown in order to match the ARPES data. The folded bands near $\Gamma$ at about $-0.2$\,eV have very little spectral weight and are omitted for clarity. \textbf{f}, \textbf{g} Constant energy maps at $E_\mathrm{F}$ at $T=17$ and 6\,K, respectively. \textbf{h}, \textbf{i} The corresponding calculated constant-energy map assuming  para- and helical magnetic phases.  \textbf{j} Calculated bands along $\Gamma$-A for a helical magnetic configuration in the vicinity of the unshifted Fermi energy. The nodes indicated by ``WP" denote the positions of the $C=\pm2$ Weyl points. \textbf{k} Hexagonal Brillouin zone of \eaa{} showing high-symmetry points. \textbf{l} Full electronic spectrum calculated with helical magnetic order. 
}
\end{figure*}



We employed spherical neutron polarimetry (SNP) to determine the magnetic structure. Eu is a very strong neutron absorber, and for an irregular-shaped crystal the absorption corrections required to determine integrated Bragg peak intensities in conventional neutron diffraction studies vary significantly for different reflections. SNP has the advantage over unpolarised ND for highly absorbing samples that the magnetic structure is determined from ratios of spin-dependent scattering intensities, and so the effect of attenuation of the diffracted neutron beam cancels~\cite{Andrews_book}. 

 Figure~\ref{fig:Figure_3}\textbf{b} plots examples of SNP data on EuAgAs recorded at $T=2$\,K at three different $(H, K, L+0.5)$ positions, with $H, K, L$ integers. For each position we plot the full polarization matrices $P_{ij}$ (comprising nine elements), where $i$ and $j$ correspond to the incident and scattered neutron polarization along the principal axes $x$, $y$, $z$ of the Blume--Maleev coordinate system~\cite{Andrews_book}. The measured polarization matrices ($P_{ij}^\mathrm{obs}$) are denoted by the grey filled circles. We find excellent agreement with the transverse helical spin structure shown in Fig.~\ref{fig:Figure_3}\textbf{a}.  We note that the $\mathbf{q}_1$ and $\mathbf{q}_2$ peaks are not resolved at 2\,K (see Fig.~\ref{fig:Figure_2}\textbf{g}), so the $P_{ij}^\mathrm{obs}$ data is an average over the two propagation vectors. The good agreement with the period-4 transverse helix model (Fig.~\ref{fig:Figure_3}\textbf{a}) suggests that the $\mathbf{q}_1$ and $\mathbf{q}_2$ structures are both close to the period-4 helix. 

Finally, Fig.~\ref{fig:Figure_REXS} shows a series of $(0,0,L)$ scans over the $(0,0,0.5)$ and $(0,0,1)$ positions in reciprocal space measured by soft x-ray REXS. Only the $\mathbf{q}_1$ peak appears in this data despite it covering a temperature range from 4 to 14\,K. We speculate that the transition signaled by the formation of the $\mathbf{q}_2$ peaks does not occur in the near surface region probed by soft x-rays. Also interesting is the observation of peaks either side of $(0,0,1)$ at $2\mathbf{q}_1$ and $(0,0,2) - 2\mathbf{q}_1$. These peaks appear below $T_{N1}$, and are about two orders of magnitude less intense than the $\mathbf{q}_1$ peak. They correspond to second-harmonic Fourier components of the $\mathbf{q}_1$ magnetic structure.



\subsection{Electronic band structure}
To understand how the Eu magnetic order affects the electronic band structure of \eaa{}, we performed angle-resolved photoemission spectroscopy (ARPES) on single-crystalline EuAgAs at various temperatures. Figure~\ref{fig:Figure_4}\textbf{a}--\textbf{c} plots the electronic dispersion along the $M$--$\Gamma$--$M$ high symmetry direction, measured at $T$= 17, 10 and 5.9\,K, respectively. These temperatures correspond to three distinct regimes discussed above: \textbf{a} the paramagnetic phase above $T_\mathrm{N1}$, \textbf{b} the intermediate phase between $T_\mathrm{N1}$ and $T_\mathrm{N2}$, and \textbf{c} the low-temperature helical magnetic phase below $T_\mathrm{N2}$.

At $T$ = 17\,K, in the paramagnetic phase,  the electronic structure displays two pairs of parallel, linearly dispersing bands [Fig.~\ref{fig:Figure_4}\textbf{a}]. The bands near the Fermi level ($E = E_\mathrm{F}$) are sharp and well-resolved. 
Interestingly, we do not find appreciable changes in the measured electronic band structure on cooling below $T_\mathrm{N1}$ to $T$ = 10\,K [Fig.~\ref{fig:Figure_4}\textbf{b}] compared to that in the paramagnetic phase. Moreover, the measured spectrum remains virtually unchanged upon further cooling below $T_\mathrm{N2}$, to $T$ = 5.9\,K [Fig.~\ref{fig:Figure_4}\textbf{c}].

Next, we examined whether the Fermi surface of \eaa{} undergoes any changes as a function of temperature across these transitions. Figure~\ref{fig:Figure_4}\textbf{f} shows the Fermi surface measured at $T$ = 17\,K, revealing two concentric circular contours centered at the $\Gamma$ point. Again, similar to the line cuts along $M$--$\Gamma$--$M$, the Fermi surface remains unchanged to within the precision of the measurement into the helical phase at $T$ = 6\,K [Fig.~\ref{fig:Figure_4}\textbf{g}].

To understand the ARPES results, we performed density functional theory (DFT) calculations for both the paramagnetic and helical magnetic phases. The calculated band dispersions are shown in Figs.~\ref{fig:Figure_4}\textbf{d} and \textbf{e}. We have performed a rigid band shift upwards of 0.5\,eV in order to match the ARPES data. The onset of helical magnetic order approximately doubles the magnetic unit cell along the $c$ axis, leading to subtle band folding near the $\Gamma$ point. However, these folded bands carry negligible spectral weight and are not visible in the experimental ARPES spectra. Similarly, the calculated (band-shifted) Fermi surfaces  for the paramagnetic and helical states [Figs.~\ref{fig:Figure_4}\textbf{h} and \textbf{i}] show minimal changes, further confirming that the helical order leaves the Fermi surface topology largely intact.

Notably, a very similar magnetic order in the structurally related compound EuCuAs has been shown to generate higher-order Weyl nodes~\cite{Soh_2024_EuCuAs} with effective Chern numbers $C$=$\pm2$ (each comprising a pair of $C$=$\pm1$ nodes so close together as to be unresolvable) in the vicinity of the unshifted Fermi surface. Our DFT results indicate that \eaa{} hosts analogous Weyl points with nontrivial Chern numbers, as highlighted in the calculated dispersion along $\Gamma$–A [Fig.~\ref{fig:Figure_4}\textbf{j}]. Whilst the ARPES measurements support the calculated band structure, direct observation of the exotic Weyl points remains challenging because they lie along directions perpendicular to the natural cleavage plane and at energies above $E_\mathrm{F}$. 

\section{Discussion}

Our combined neutron diffraction, resonant elastic x-ray scattering, spherical neutron polarimetry, ARPES, and first-principles calculations establish \eaa{} as a rare example of a hexagonal pnictide that hosts long-range helical magnetic order coexisting with robust topological electronic states. The observation of two successive magnetic transitions, culminating in a non-collinear helical configuration that doubles the unit cell along the $c$ axis, demonstrates the rich magnetic phase space accessible in this system.

This study has shown that the magnetic structure of \eaa{} in the temperature range $T_\mathrm{N2}<T<T_\mathrm{N1}$ is a transverse helix in which the Eu moments lie in the $ab$ plane and rotate by an angle $\phi = \pi q_1 \simeq \pi/2$, where $q_1$ is in r.l.u., between adjacent Eu layers ($\phi$ is in fact slightly greater than $\pi/2$ and temperature dependent, see Fig.~\ref{fig:Figure_2}\textbf{c}). The presence of weak second harmonics suggests that the helix may not be precisely circular.  The emergence of the $\mathbf{q}_2$ peaks below $T_\mathrm{N2}$ signifies a second transverse helical magnetic component, with a turn angle $\phi$ of slightly less than $\pi/2$. 
The very small heat capacity anomaly observed at $T_\mathrm{N2}$ compared with the large and sharp anomaly at $T_\mathrm{N1}$ \cite{EuAgAs_DFT_Magnetization} implies that there is little difference in entropy between the two magnetic phases, indicating that a subtle modification of the high temperature helical structure occurs below $T_\mathrm{N2}$, perhaps due to frustration or details of the Fermi surface which may influence metallic exchange.  When both $\mathbf{q}_1$ and $\mathbf{q}_2$ components are present, the transverse helix acquires a long-range modulation along the $c$ axis with a period of $[2\pi(q_1-q_2)]^{-1}c \sim 8c$.

Our ARPES results show that the Fermi surface remains largely unchanged despite the onset of helical order, in agreement with DFT predictions after a rigid band shift. The symmetries that are broken in the magnetically ordered phase lead to subtle band folding and splitting at energies near the calculated Fermi energy. In particular, the breaking of both inversion and time-reversal symmetry that protects the two-fold degeneracy of the bands in the paramagnetic phase leads to the emergence of Weyl nodes at the A points of the Brillouin zone, Fig.~\ref{fig:Figure_4}\textbf{j,k}. This behavior parallels that of EuCuAs~\cite{Soh_2024_EuCuAs}, reinforcing the notion that such hexagonal rare-earth pnictides provide a fertile platform for realizing magnetic Weyl semimetals driven by non-collinear spin textures.

Our findings highlight \eaa{} as a promising candidate for studying the interplay between helical magnetism and electronic band topology. The tunable nature of its magnetic order may enable future control of Weyl node properties by external fields, pressure, or chemical substitution. Further high-resolution ARPES measurements of 100 and 110 cleavage surfaces, combined with transport studies and scanning probe experiments, will be essential to directly detect the Weyl nodes and explore emergent phenomena such as the topological Hall effect or chiral anomaly in this material class.

In summary, \eaa{} expands the family of magnetic topological semimetals and provides a model system for exploring novel quasiparticles and correlated topological phases in non-collinear magnetic backgrounds. 

\textbf{Note added:} During preparation of this manuscript, we became aware of a related preprint~\cite{gazzah2026} reporting unpolarized neutron diffraction measurements on EuAgAs, which identified magnetic ordering with propagation vector $\mathbf{q}$=$(0,0,0.5)$. In contrast, our measurements reveal two temperature-dependent incommensurate propagation vectors, $\mathbf{q}_1$=$(0,0,0.5+\delta_1)$ and $\mathbf{q}_2$=$(0,0,0.5-\delta_2)$, observed below $T_\mathrm{N1}$ and $T_\mathrm{N2}$, respectively. Furthermore, our polarized neutron measurement enables discrimination between helical and collinear magnetic structures, which may not be distinguishable by unpolarized neutron diffraction alone due to domain-averaging effects~\cite{Soh_2024_EuCuAs}. The discrepancy in propagation vectors may reflect sample-dependent effects, including possible variations in stoichiometry.

\section{Methods}
\textbf{Crystal growth and characterization} Single crystalline \eaa\, was grown by the self-flux method, as described in Ref.~\cite{EuAuAs_DFT_Magnetization}. The structural quality of the crystals was verified with laboratory x-rays on a 6-circle diffractometer (Oxford Diffraction) and a Laue diffractometer (Photonic Science). The x-ray data confirmed the reported $(P6_3/mmc)$ structure and revealed a very sharp  mosaic. Several crystals from the same batch were selected for the measurements. Magnetization measurements were performed on a crystal of mass 6.5\,g with a superconducting quantum interference device (SQUID) magnetometer (MPMS-7, Quantum Design). The temperature-dependent magnetometry measurements were performed in the temperature range $2 \leq T \leq 30$\,K in an external field of $H = 1000$\,Oe applied parallel to the hexagonal $a$ axis.

\textbf{Single-crystal magnetic neutron diffraction.} Single-crystal neutron diffraction measurements were performed on a crystal of mass 103\,mg on the D9 four-circle diffractometer at the Institut Laue–Langevin (ILL). To reduce the strong neutron absorption of Eu, a hot neutron beam with a wavelength of 0.84\,{\AA} was employed. An erbium filter was installed in the incident beam to suppress higher-order wavelength contamination.

\textbf{Resonant elastic x-ray diffraction.} Soft REXS measurements were carried out at the UE46 PGM-1 beamline at BESSY II. The incident X-ray photon energy was tuned to the Eu $M_5$ absorption edge to enhance the scattering signal from the Eu$^{2+}$ ions through resonant magnetic scattering. The diffractometer was configured in a horizontal scattering geometry, with $\pi$-polarized incident photons to probe magnetic order via rotation of the linear polarization into the $\pi \rightarrow \sigma^\prime$ channel, which is predominantly sensitive to magnetic scattering.

\textbf{Polarized neutron diffraction.} Polarized neutrons with a wavelength of 0.83\,\AA{} were employed on the D3 diffractometer at the Institut Laue-Langevin (ILL) using a spherical neutron polarimetry (SNP) setup equipped with a Cryopad device~\cite{Tasset1999,Lelievre2005}. The crystal was aligned with the hexagonal $b$ axis vertical so as to access reflections in the $(H,0,L)$ plane in reciprocal space. The incident beam was polarized by Bragg reflection from the (111) planes of a ferromagnetic Heusler alloy crystal (Cu$_2$MnAl), and an erbium filter was inserted in the incident path to suppress half-wavelength contamination. Nutator and precession fields controlled the polarization direction of the incident beam. A $^{3}$He spin filter was used to analyse the polarization of the scattered neutrons. Standard corrections for the time-dependent decay of the spin filter efficiency were applied using repeated measurements of the (102) structural Bragg peak. The polarization matrix elements $P_{ij}$ are defined as
\begin{align}
P_{ij} = \frac{N_{ij} - N_{i\bar{j}}}{N_{ij} + N_{i\bar{j}}},
\nonumber
\end{align}
where $N_{ij}$ and $N_{i\bar{j}}$ denote the counts at a given Bragg reflection when the incident neutron polarization is along direction $i$ and the scattered polarization is measured parallel or antiparallel to $j$, respectively. Here, $x$ is defined along the scattering vector $\textbf{\textit{Q}}$, $z$ is perpendicular to the scattering plane, and $y$ completes the right-handed coordinate system. SNP measurements on EuAgAs were made at ten magnetic peak positions, and in two cases the measurement was repeated with the opposite incident neutron polarization.
 
\textbf{Density functional theory.} To elucidate the topological character of the electronic band structure in \eaa{}, we performed density functional theory (DFT) calculations using VASP~\cite{VASP1,VASP2} version 6.2.1. The calculations employed projector-augmented wave (PAW) pseudopotentials within the generalized gradient approximation (GGA) parameterized by Perdew, Burke, and Ernzerhof (PBE)\cite{Perdew96}. To capture the significant spin–orbit coupling (SOC) effects arising from the heavy As ions—which can induce band inversion—relativistic pseudopotentials were used, with a kinetic energy cutoff set to 480,eV. Strong electron–electron correlations in the localized Eu $4f$ states were modeled using a Hubbard $U$ of 5.0,eV, reproducing the experimentally observed binding energy (see Supplementary Information). The Brillouin zone was sampled using a $9 \times 9 \times 7$ Monkhorst–Pack \textbf{k}-point mesh\cite{PhysRevB.13.5188}.

\textbf{Angle-resolved photoemission spectroscopy.} ARPES experiments were carried out at high-resolution branch (HR-ARPES)~\cite{Hoesch_RSI_2017} of the beamline I05 in the Diamond Light Source, equipped with a MBS A-1 analyser. Samples were cleaved in situ at approximately 6\,K under ultra-high vacuum conditions better than 2$\times$10$^{-10}$ mbar and measured at temperatures ranging from 6\,K to 17\,K using an incident photon energy of 85\,eV. To locate the $\Gamma$ and A high-symmetry points, we performed a $k_z$-dependent scan by tuning the photon energy between 50\,eV and 160\,eV. Measurements were conducted with horizontally polarized light.

\section{Data availability}
{\color{black} The neutron scattering data presented in this paper are available at~\cite{NDD9} (D9, ILL, proposal 5-41-1261) and ~\cite{SNPD3} (D3, ILL, proposal DIR-295). The rest of the data, including resistivity, magnetization and REXS, will be made available on the Zenodo repository. }

\bibliographystyle{unsrt}
\bibliography{ref12.bib}

\section{Acknowledgments}
\begin{acknowledgments}
The authors  wish  to  thank Yang Xupeng for extensive help with the experiments. The proposal numbers for the data presented in this manuscript are 5-41-1261 (D9, ILL~\cite{NDD9}) and DIR-295 (D3, ILL~\cite{SNPD3}). D.P. and A.T.B. acknowledge support from the Oxford–ShanghaiTech collaboration project. This work was supported by the U.K. Engineering and Physical Sciences Research Council, grant no.~EP/M020517/1. J.-R.S. acknowledges support from the Singapore National Science Scholarship, Agency for Science Technology and Research and the European Research Council (HERO, Grant No. 810451). 
\end{acknowledgments}

\section{Author contributions}
J.-R.S. and A.T.B. conceived the experiments. D.P. grew the single crystals, and J.-R.S., D.P., L.W. and A.T.B. characterised and performed bulk measurements on the crystals. Unpolarized neutron diffraction was carried out by A.T.B., J.A.R.-V. and O.F. The REXS experiment was conducted by J.-R.S., and E. W.. T. K. and D. Y. performed the ARPES experiment. The SNP experiment was performed by J.-R.S., A.T.B., J.A.R.-V. and A.S.. The \textit{ab initio} electronic structure calculations and interpretation were performed by Y.S.Y and Z.M.. All authors reviewed the manuscript.

\section{Competing interests}
The authors declare no competing financial or non-financial interests.

\end{document}